\documentclass[reprint,aps,prl,amsmath,amssymb,nofootinbib,superscriptaddress, preprintnumbers]{revtex4-2}

\usepackage[utf8]{inputenc}
\usepackage[T1]{fontenc}

\usepackage{amsmath}
\usepackage{mathrsfs}
\usepackage{txfonts}
\usepackage{mathtools}
\usepackage{braket}
\usepackage{tensor}
\usepackage{xcolor}
\usepackage[abbreviations]{siunitx}
\usepackage{graphicx}
\usepackage{booktabs}

\usepackage{float}

\usepackage{braket}
\usepackage{ulem}

\usepackage{hyperref}
\usepackage{cleveref}

\newcommand{\beq}{\begin{equation}}
\newcommand{\eeq}{\end{equation}}
\newcommand{\beqn}{\begin{eqnarray}}
\newcommand{\eeqn}{\end{eqnarray}}
\newcommand{\bsub}{\begin{subequations}}
\newcommand{\esub}{\end{subequations}}
\newcommand{\bpm}{\begin{pmatrix}}
\newcommand{\epm}{\end{pmatrix}}

\newcommand{\ndnd}{$^{150}$Nd+$^{150}$Nd}
\newcommand{\smsm}{$^{150}$Sm+$^{150}$Sm}

\newcommand{\zrzr}{$^{96}$Zr+$^{96}$Zr}
\newcommand{\ruru}{$^{96}$Ru+$^{96}$Ru}

\newcommand{\auau}{$^{197}$Au+$^{197}$Au}
\newcommand{\uuuu}{$^{238}$U+$^{238}$U}

\newcommand{\trento}{T$\mathrel{\protect\raisebox{-0.5ex}{R}}$ENTo}

\begin{document}
    
\title{Benchmarking nuclear matrix elements of  $0\nu\beta\beta$ decay  with high-energy nuclear collisions}

  \author{Yi Li}   
  \affiliation{School of Physics and Astronomy, Sun Yat-sen University, Zhuhai 519082, P.R. China}   
 \affiliation{Guangdong Provincial Key Laboratory of Quantum Metrology and Sensing, Sun Yat-Sen University, Zhuhai 519082, China } 

  \author{Xin Zhang}   
    \affiliation{Department of Physics, Kyoto University,
Kyoto 606-8502, Japan}    
    
    \author{Giuliano Giacalone} 
    \email{giuliano.giacalone@cern.ch}
  \affiliation{Theoretical Physics Department, CERN, CH-1211 Gen\`eve 23, Switzerland}    
  
  \author{Jiangming Yao}    
  \email{yaojm8@sysu.edu.cn}  
  \affiliation{School of Physics and Astronomy, Sun Yat-sen University, Zhuhai 519082, P.R. China}    
   \affiliation{Guangdong Provincial Key Laboratory of Quantum Metrology and Sensing, Sun Yat-Sen University, Zhuhai 519082, China }

\date{\today}

\begin{abstract}
Reducing uncertainties in the nuclear matrix elements (NMEs) remains a critical challenge in designing and interpreting experiments aimed at discovering  neutrinoless double beta ($0\nu\beta\beta$) decay. Here, we identify a class of observables, distinct from those employed in low-energy nuclear structure applications, that are strongly correlated with the NMEs: momentum correlations among hadrons produced in high-energy nuclear collisions. Focusing on the $^{150}$Nd$\rightarrow$$^{150}$Sm transition, we combine a Bayesian analysis of the structure of $^{150}$Nd with simulations of high-energy \ndnd{} collisions. We reveal prominent correlations between the NMEs and features of the quark-gluon plasma (QGP) formed in these processes, such as spatial gradients and anisotropies, which are accessible via collective flow measurements. Our findings demonstrate collider experiments involving $0\nu\beta\beta$ decay candidates as a platform for benchmarking theoretical predictions of the NMEs.
\end{abstract}

\preprint{CERN-TH-2025-035}

\maketitle

\begin{figure}[t]
    \centering
    \includegraphics[width=.99\linewidth]{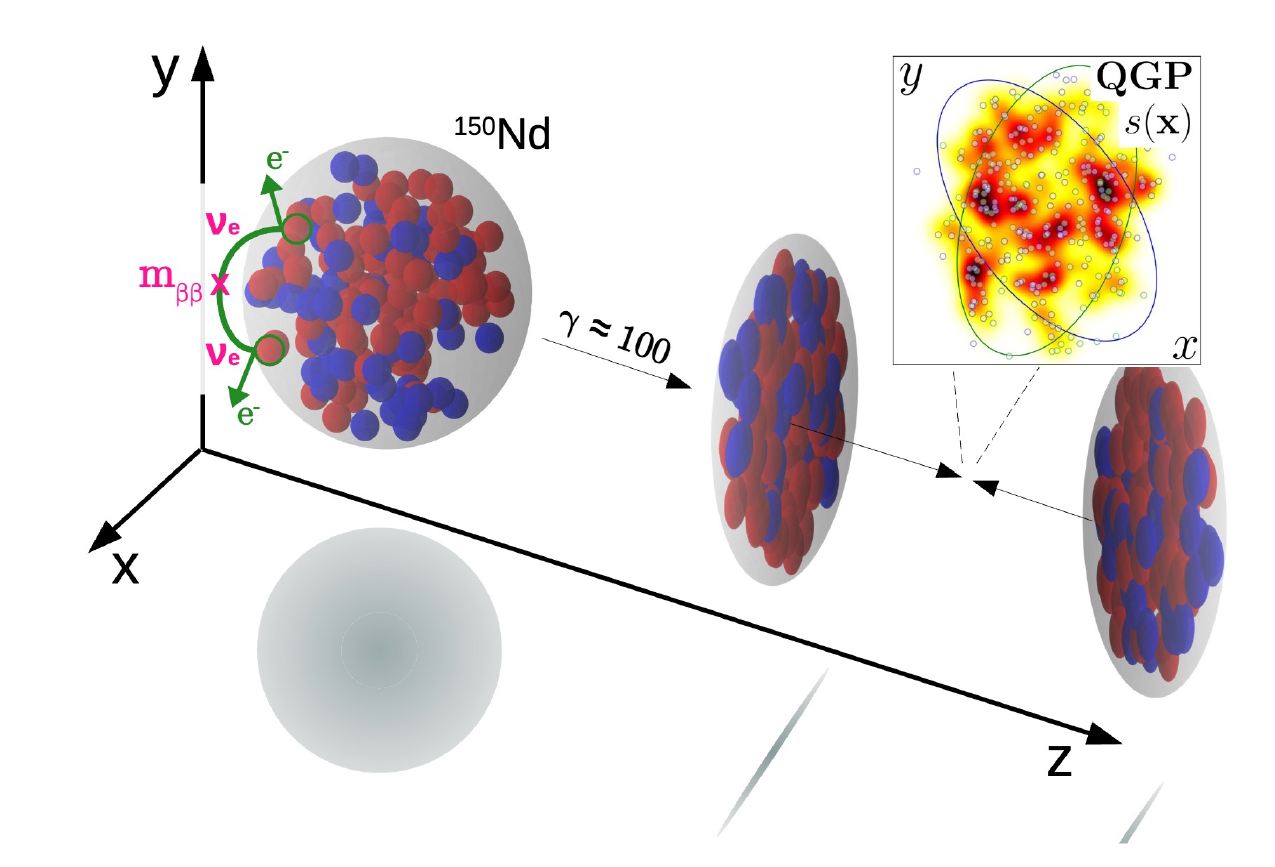}
    \caption{\textit{Left:} The nucleus \nuclide[150]{Nd} may undergo $0\nu\beta\beta$ decay, with two neutrons converting into protons via the emission of two electrons and the exchange of a light electron neutrino with Majorana mass $m_{\beta\beta}$. The NME of the transition is highly sensitive to the geometry of the nucleon distribution.  \textit{Right:} Two \nuclide[150]{Nd} nuclei boosted at ultrarelativistic speed collide at zero impact parameter, forming a QGP whose density profile retains information about the spatial arrangement of the interacting nucleons, opening thus a new experimental window onto nuclear properties correlated with the NME. The Lorentz factor  is defined as $\gamma = 1/\sqrt{1 - v_z^2/c^2}$.
    }
    \label{fig:cartooon}
\end{figure}

 
\textit{\textbf{Introduction.}}  
The discovery of neutrino oscillations shows that neutrinos are massive particles~\cite{SNO:2001,KamLAND:2003,DayaBay:2012}, providing a compelling case for the search of $0\nu\beta\beta$ decay, a hypothetical nuclear transition~\cite{Furry:1939} in which two neutrons decay into two protons with the emission of two electrons, but no (anti)neutrinos. Its observation would allow us to determine the nature of the neutrino (Dirac or Majorana)~\cite{Schechter:1982}, and would demonstrate the violation of the lepton number in nature, shedding light on the origin of the matter-antimatter asymmetry~\cite{Fukugita:1986}. Furthermore, if this process is driven by the standard mechanism of light Majorana neutrino exchange (left of Fig.~\ref{fig:cartooon}), the half-life of $0\nu\beta\beta$ decay could give insights into the absolute neutrino masses and their hierarchy. Achieving reliable theoretical control over the NMEs for $0\nu\beta\beta$ decay in various candidate nuclei—which connect experimental signals to the effctive neutrino mass—is essential for both the design and interpretation of current and future ton-scale experiments~\cite{Agostini:2023}.

Obtaining knowledge of the NMEs for different candidate nuclei that is both accurate and consistent across models poses a major challenge for nuclear theory. The values predicted by different models vary by factors of three or more, leading to an uncertainty of an order of magnitude in the half-life for a given effective neutrino mass~\cite{Engel:2017,Yao:2022PPNP,Agostini:2023}. Despite significant progress in calculating the NMEs of candidate nuclei from first principles~\cite{Yao:2020PRL,Belley:2021,Novario:2021,Belley:2024zvt}, the uncertainty related to nuclear interactions and many-body truncations remains substantial~\cite{Belley:2024PRL}. To mitigate this problem, considerable effort has been devoted to exploring correlations between the NMEs and experimental observables from low-energy nuclear structure experiments, such as double Gamow-Teller transitions~\cite{Shimizu:2018,Yao:2022PRC,Jokiniemi:2023,Wang:2024} and double-gamma transitions~\cite{Romeo:2022,Romeo:2025}. Notably, correlations between the NMEs and the properties of low-lying states of the initial and final nuclei have been examined using a Bayesian analysis within the interacting shell models~\cite{Horoi:2022,Horoi:2023}, the valence-space in-medium similarity renormalization group~\cite{Belley:2022_correlation}, as well as the multi-reference covariant density functional theory (MR-CDFT)~\cite{Zhang:2024_short,Zhang:2024_long}. These studies have revealed strong correlations between the NMEs, the excitation energies of the first $2^+$ states, and the $B(E2;0^+_1\to 2^+_1)$ values of the candidates. These findings corroborate the observation that the NMEs are sensitive to the deformation parameters of the candidate isotopes~\cite{Rodriguez:2010PRL,Yao:2015_NLDBD,Belley:2024PRL,Jiao:2023}.

At the other end of the energy spectrum, collider studies have established that measurements of the collective flow of hadrons in the soft sector of high-energy nuclear collisions enable us to experimentally access fine properties of the shapes and radial distributions of nuclei in their ground states~\cite{STAR:2015mki,ALICE:2018lao,CMS:2019cyz,ATLAS:2019dct,STAR:2021mii,ALICE:2021gxt,ATLAS:2022dov,STAR:2024wgy,ALICE:2024nqd}. In particular, by studying how observables such as anisotropic flow coefficients vary across collision systems involving isobaric isotopes \cite{Giacalone:2021uhj}, one can measure signatures of the structure of these nuclei while drastically mitigating the impact of theoretical uncertainties on poorly understood features of the QGP, such as out-of-equilibrium or hadronization phenomena, in the interpretation of the data \cite{Xu:2021uar,Nijs:2021kvn,Zhang:2022fou,Zhao:2022uhl,Gardim:2023ksn,Giacalone:2024luz,Giacalone:2024ixe,Xu:2024bdh,Mantysaari:2024uwn}.  Therefore, collider experiments with isobars offer robust probes of nuclear geometry. Since $0\nu\beta\beta$ decay is expected to occur between two isobaric ground states, it is natural to leverage this information to evaluate how high-energy experiments can constrain theoretical determinations of the NMEs.

In this Letter, we take a first step in this direction. We couple the aforementioned MR-CDFT-based Bayesian analysis \cite{Zhang:2024_short,Zhang:2024_long} to state-of-the-art simulations of high-energy nuclear collisions, and study the correlation between the NME of the $^{150}$Nd$\rightarrow$$^{150}$Sm transition and geometric properties of the QGP that are experimentally accessible via collective flow measurements. The signal we find is as strong as that obtained by correlating the NME with more traditional electromagnetic probes of the nuclear geometry. 
We demonstrate, thus, an experimental technique for the benchmark of NME calculations that is complementary to low-energy nuclear physics methods.


\textit{\textbf{NME, nuclear structure, and heavy-ion collisions.}} The NME of $0\nu\beta\beta$ decay for the transition from the ground state ($0^+_1$) of \nuclide[150]{Nd} to the ground state ($0^+_1$) of \nuclide[150]{Sm} is given by
\beq 
M^{0\nu} = \bra{\Psi_F(0^+_1)}\hat{O}^{0\nu}\ket{\Psi_I(0^+_1)},
\eeq
where $\hat{O}^{0\nu}$ is the transition operator based on the standard mechanism of exchange of light Majorana neutrinos.  The decay operator includes contributions from vector coupling (VV), axial-vector coupling (AA), interference of axial-vector and induced pseudoscalar coupling (AP), induced pseudoscalar coupling (PP), and weak-magnetism coupling (MM) terms, which are related to the products of two current operators~\cite{Song:2014,Yao:2015_NLDBD}; see the Supplemental Material~\cite{SupplementalMaterial}. We note that the contact transition operator~\cite{Cirigliano:2018} is not included in this evaluation, as it has been shown that, within the relativistic framework, the renormalizability of the transition amplitude is automatically ensured at leading order~\cite{Yang:2024PLB}, rendering the contact transition operator a subleading contribution. Furthermore, we neglect the effects of two-body currents, which appear at next-to-next-to-next-to-leading order in chiral effective field theory, and which are known to reduce the nuclear matrix elements by approximately 10\%~\cite{Wang:2018_TBC}.

The ground-state wave functions, $\ket{\Psi_{I/F}(0^+)}$, of the candidates are determined within the MR-CDFT approach. As detailed in the Supplemental Material~\cite{SupplementalMaterial}, the wave functions are constructed as linear combinations of particle-number ($N$, $Z$), and angular-momentum ($J$) projected mean-field wave functions, $\ket{\Phi(\mathbf{q})}$, with collective coordinate $\mathbf{q}$,
\begin{equation}
\label{eq:gcm_wf}
\ket{\Psi_{I/F}(J^+_\nu)}
=\sum_{\mathbf{q}} f^{JNZ}_\nu(\mathbf{q}) \hat P^{N}\hat P^{Z}\hat P^{J}\ket{\Phi(\mathbf{q})},
\end{equation}
where  $f^{JNZ}_{\nu}({\bf q})$  is a weight, and $\hat P$ are projection operators. The states $\ket{\Phi(\mathbf{q})}$ are determined through a deformation-constrained relativistic mean-field (RMF) model combined with  Bardeen–Cooper–Schrieffer (BCS) theory~\cite{Burvenich:2001rh,Zhao:2010}. For the nuclear Hamiltonian, we employ a relativistic energy density functional (EDF) composed of the kinetic energy, $\tau(\boldsymbol{r})$, the electromagnetic energy, $\mathcal{E}^{\text {em}}(\boldsymbol{r})$, as well as the nucleon-nucleon ($NN$) interaction energy~\cite{Burvenich:2001rh,Zhao:2010,Zhang:2024_long},
 \begin{equation}
 \label{eq:EDF}
    E[\tau, \rho,\nabla\rho; \mathbf{C}] 
    =\int d^3r \Big[\tau(\boldsymbol{r})
    +\mathcal{E}^{\text {em}}(\boldsymbol{r})
    + \sum^9_{\ell=1} c_\ell  \mathcal{E}^{\rm NN}_\ell(\boldsymbol{r}; \rho, \nabla \rho) \Big],
\end{equation}
where $\rho$ and $\nabla \rho$ represent the scalar, vector and isovector densities (currents), and their derivatives. The $NN$ interaction energy, $\mathcal{E}^{\rm NN}$, is parameterized as a function of $\rho$ via parameters $c_\ell$ collectively denoted by $\mathbf{C}=\{\alpha_S, \beta_S, \gamma_S, \delta_S, \alpha_V, \gamma_V, \delta_V, \alpha_{TV}, \delta_{TV}\}$. The subscripts $(S, V)$ indicate the scalar and vector types of $NN$ interaction vertices in Minkowski space, respectively, while subscript $T$ is for the vector type in isospin space \cite{Zhao:2010}.

Subsequently, the ground-state wave functions obtained from Eq.~(\ref{eq:gcm_wf}) are used to initialize simulations of high-energy \ndnd{} collisions. With knowledge of the first $0^+$ and $2^+$ states, we compute the electric quadrupole transition strength $B(E2; 0^+\to2^+)$. As both \nuclide[150]{Nd} and \nuclide[150]{Sm} are well-deformed nuclei \cite{Yao:2014uta}, it is reasonable to relate the quadrupole deformation parameter, $\beta_2$, of the intrinsic proton density to the square root of $B(E2)$~\cite{RAMAN20011}:
 \begin{equation}  
 \label{eq:beta2}
    \beta_2 = \frac{4\pi}{3ZR_0^2} \sqrt{B(E2; 0^+_1\to2^+_1)},
\end{equation}
where $R_0=1.2A^{1/3}$ fm. Due to the ultra-short time scales involved, a heavy-ion collision at high energy is understood to take a snapshot of the positions of the nucleons as sampled from the nuclear intrinsic shape with a random orientation in space (right of Fig.~\ref{fig:cartooon}). Following  previous studies \cite{Bally:2021qys,Bally:2023dxi,Ryssens:2023fkv,Giacalone:2024luz}, the relevant intrinsic density in three-dimensional space, $\rho_V({\bf r})$, is determined through a RMF+BCS calculation, denoted single-reference (SR)-CDFT, with a constraint on the mass quadrupole deformation parameter, $\bar \beta_2=\beta_2$.  Using $\rho_V({\bf r})$ with the constrained $\bar \beta_2$ as input, we simulate \ndnd{} collisions at ultra-relativistic energy.

The simulations follow the \trento{} model of initial conditions \cite{Moreland:2014oya}, whose working principles are recalled in the Supplemental Material~\cite{SupplementalMaterial}. The output of the collisions is the entropy density of the QGP in the transverse plane, $s({\bf x})$, one example of which is shown in Fig.~\ref{fig:cartooon}. To relate this density to experimental observables, two quantities are used:
\begin{align}
\nonumber  \varepsilon_2 &= |\mathcal{E}_2|,  \hspace{20pt} {\mathcal E}_2 = -\frac{\int_{\bf x}  (x+i y)^2 s(\mathbf{x})}{ \int_{\bf x}   (x^2 + y^2) s(\mathbf{x})}, \\
 E/S &= \frac{\int_{\bf x} e({\bf x})}{\int_{\bf x} s({\bf x})} , \   \hspace{20pt}  e({\bf x}) \propto s({\bf x})^{4/3}, 
\end{align}
where ${\bf x}=(x,y)$. The first quantity, $\varepsilon_2$, is the dimensionless spatial ellipticity of the QGP \cite{Teaney:2010vd}, strongly sensitive to the quadrupole deformation of the colliding nuclei \cite{Giacalone:2021udy}. The quantity $E/S$ is proportional to the ratio of the total energy of the QGP to its total entropy. The energy density, $e({\bf x})$, is obtained by applying the QCD equation of state. In particular, at the initial condition of heavy-ion collisions the temperature of the QGP is high enough to justify a simple ideal gas prescription, namely, $e({\bf x}) \propto s({\bf x})^{1+c_s^2}$, with $c_s^2 = 1/3$. The total energy is determined, thus, by the steepness of the spatial gradients of the entropy density. As $S$ determines the amount of final-state hadrons, $E/S$ is proportional to the energy per particle.  

We simulate 10$^6$ collisions (or \textit{events}) at zero impact parameter (ultra-central collisions), which present a strong sensitivity to nuclear deformations. From the ensemble of $\varepsilon_n$ and $E/S$ values, we evaluate the following averages:
\begin{equation}
\label{eq:obsEn}
  \left   \langle \varepsilon_2^2 \right  \rangle, \hspace{20pt} \left \langle ( \delta E/S )^2 \right \rangle, \hspace{20pt} \left \langle \varepsilon_2^2 \, \delta E/S \right \rangle, \hspace{20pt}  2  \left \langle \varepsilon_2^2  \right\rangle^2 -  \left \langle \varepsilon_2^4 \right \rangle  
\end{equation}
with $\delta E/S = E/S - \langle E/S \rangle$, which give access to final-state observable quantities, as we explain in the next section.

\begin{figure}[bt]
    \centering
    \includegraphics[width=0.8\linewidth]{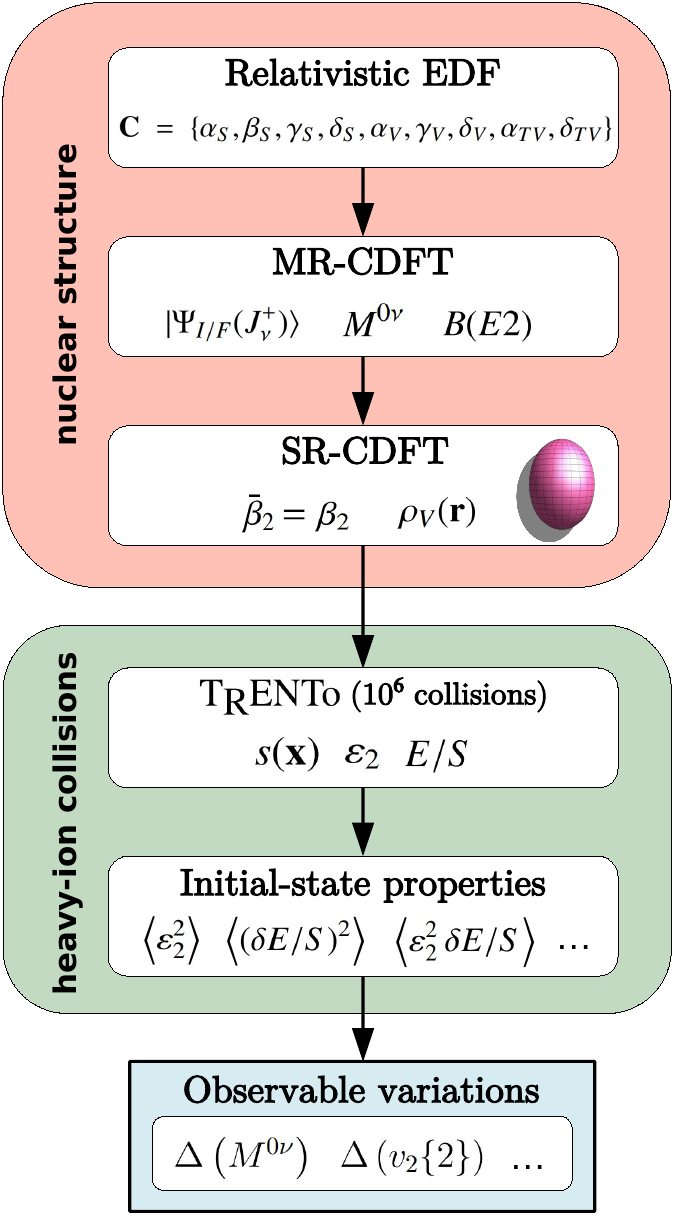}
    \caption{Flow chart of our framework that combines the results of a Bayesian analysis of low-energy nuclear structure data in the CDFT framework with high-energy heavy-ion collision simulations.}
    \label{fig:flowc}
\end{figure}

Our workflow is summarized in Fig.~\ref{fig:flowc}. We start by generating a set, ${\bf C}$, of parameters of the nuclear EDF. From that, we evaluate the NME, $M^{0\nu}$, as well as an intrinsic nuclear shape with quadrupole deformation, $\beta_2$, determined from the transition strength $B(E2)$. The intrinsic nuclear shape is used to simulate 10$^6$ ultra-central \ndnd{} collisions. We obtain thus a distribution of $\varepsilon_n$ and $E/S$ values, from which we evaluate the averages in Eq.~(\ref{eq:obsEn}). The process is repeated for 10$^3$ samples of the EDF parameters, which are obtained by varying the nine parameters $\mathbf{C}$ around their optimal values of the PC-PK1~\cite{Zhao:2010} using quasi Monte-Carlo  sampling with a uniform distribution~\cite{Zhang:2024_long}. We emphasize that each initial choice of ${\bf C}$ provides a reasonable description of both nuclear matter properties and the low-energy structure of \nuclide[150]{Nd}\cite{Zhang:2024_long}. Consequently, the spread in the results for the quantities in Eq.(\ref{eq:obsEn}) should be interpreted as the theoretical uncertainty arising from the limited knowledge of nuclear deformation within a model constrained by low-energy data.

\begin{figure}[t]
 \centering 
\includegraphics[width=.98\linewidth]{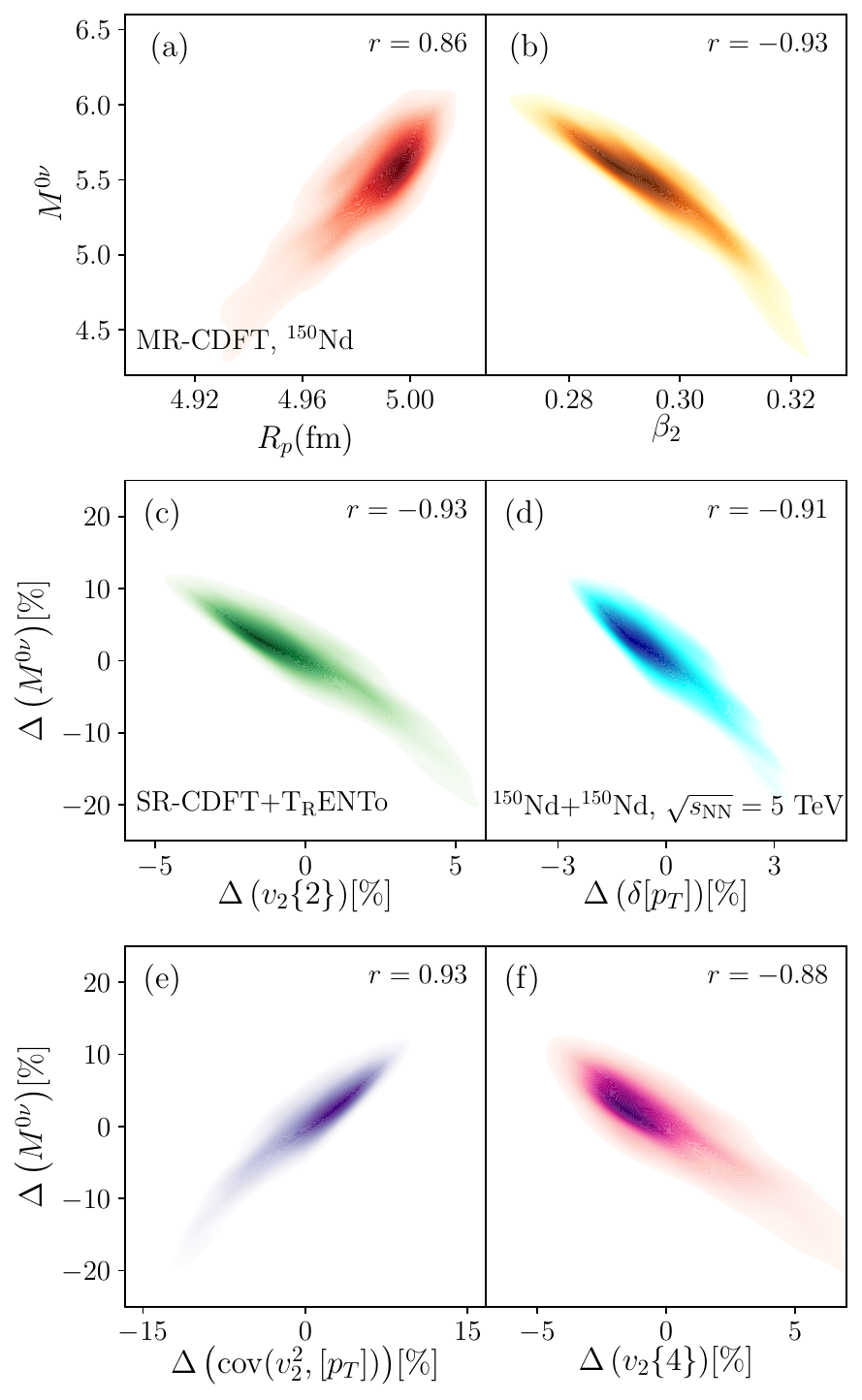}  
\caption{\textit{Panels (a-b):} Correlation between the NME of the $0\nu\beta\beta$ decay of the nucleus $^{150}$Nd, $M^{0\nu}$, and \textit{(a)} the proton radius, $R_p$, and \textit{(b)} the quadrupole deformation parameter, defined via Eq.~(\ref{eq:beta2}). \textit{Panels (c-f):} Correlation between the relative variation of the NME and the relative variations of observables accessible in ultra-central \ndnd{} collisions. \textit{(c)} Correlation with the elliptic flow, $v_2$, \textit{(d)} correlation with the fluctuation of the mean transverse momentum, $\delta [p_T]$, \textit{(d)} correlation with the covariance between $v_2^2$ and $\delta [p_T]$, and \textit{(f)} correlation with the fourth-order cumulant of the elliptic flow vector. Pearson correlation coefficients, $r$, are displayed in each panel.}.  
\label{fig:correlation_HIC_observables}
\end{figure}

\textit{\textbf{Results and discussion.}} 
Figure~\ref{fig:correlation_HIC_observables}(a)-(b) shows how the NME is correlated with the nuclear proton radius, $R_p$, and the quadrupole deformation parameter, $\beta_2$, in the Bayesian analysis. One can see that the NME is anti-correlated with the quadrupole deformation $\beta_2$ of the initial nucleus \nuclide[150]{Nd}. This is consistent with the finding of previous studies, which have further demonstrated that the NME is significantly quenched when the quadrupole deformations of the initial and final nuclei differ~\cite{Chaturvedi:2008,Fang:2010,Mustonen:2013,Rodriguez:2010PRL,Sahu:2014,Song:2014,Yao:2015_NLDBD,Fang:2018}. Here, we find that the Pearson correlation coefficient between the NME and the difference in quadrupole deformations is $r = -0.78$, which interestingly is smaller in magnitude than the value $-0.93$ for the correlation between the NME and the deformation of the parent nucleus, \nuclide[150]{Nd}. This is however slightly larger than the value $r=-0.76$ for the correlation between the NME and the deformation of \nuclide[150]{Sm}; see the Supplemental Material~\cite{SupplementalMaterial} for additional figures and discussions. 

The notable feature is that, while the relative systematic uncertainty on the proton radius is less than 0.5\%, the distribution of $\beta_2$ shows variations up to 10\% around the mean value. 
This indicates that fitting observables such as B(E2) leads to an imprecise determination of the nuclear matrix element (NME), with a relative systematic uncertainty exceeding 10\%. Although data from low-energy nuclear structure experiments provide essential information about the electromagnetic properties of low-lying nuclear states, they appear insufficient to strongly constrain the many-body structure of the ground states, which is instead directly accessed through multi-particle correlation measurements at colliders \cite{Giacalone:2023hwk}. Moreover, low-energy experiments do not yield direct information about the deformation of bulk nuclear matter; such information can in principle be inferred from inelastic scattering experiments, whose interpretation is however subject to significant model dependence. This highlights the central message of this Letter: the complementarity of low- and high-energy approaches should be exploited to simultaneously capture the electromagnetic and excited-state properties of nuclei (from low-energy experiments) and the full complexity of their ground states (from high-energy experiments). Doing so will lead to a more complete understanding of the many-body nuclear wave function and more reliable NME evaluations.

With this in mind, we now move on to discuss correlations between the NME and observable features of the QGP produced in \ndnd{} collisions. We recall that signatures of nuclear structure are mainly observed in the distribution of the soft hadrons emitted to the final states \cite{Giacalone:2023hwk}. Their spectrum in a class of events (e.g., ultra-central collisions) in the transverse plane at $z=0$ (following, e.g., Fig.~\ref{fig:cartooon}) is measured differentially in transverse momentum, ${\bf p}=(p_T,\phi)$. Its angular distribution is decomposed in modes \cite{Ollitrault:2023wjk}:
\begin{equation}
\frac{dN_{\rm ch}}{d^2 {\bf p} } \propto \frac{dN_{\rm ch}}{p_T d p_T}   \biggl ( 1 + \sum_{n>1} 2 \, v_n \cos n(\phi-\phi_n) \biggr ) \, .    
\end{equation}
In what follows we shall need the second  harmonic, $v_2$, which quantifies the \textit{elliptic flow} of the system, as well as the average hadron momentum, $[p_T]= \frac{1}{N_{\rm ch}} \int_{\bf p} p_T \frac{dN}{d^2 {\bf p}}$. 

 Hydrodynamic simulations indicate that the final-state elliptic flow coefficient, $v_2$, is in a strong linear correlation with the initial-state ellipticity parameter, $\varepsilon_2$ \cite{Niemi:2012aj,Noronha-Hostler:2015dbi,Sousa:2024msh}. Similarly, the average transverse momentum (which measures the energy per particle) is correlated with the value of $E/S$ \cite{Schenke:2020uqq,Giacalone:2020dln}. For an observable $\mathcal{O}$, we define its relative variation by:
\begin{equation}
    \Delta (\mathcal{O}) = \frac{\mathcal{O}-\langle \mathcal{O} \rangle_{\bf C}}{ | \langle \mathcal{O} \rangle_{\bf C} | },
\end{equation}
where the subscript ${\bf C}$ denotes an average over samples of the EDF parameter set.  We consider, then, the variation of the following observables, where the pre-factors ensure that they can be meaningfully compared to each other:
\begin{align}
\label{eq:obsVn}
\nonumber \Delta \left ( v_2 \{ 2 \}\right )  &\equiv  \frac{1}{2} \Delta \left( \left\langle  v_2^2 \right\rangle   \right), \\
\nonumber \Delta \left ( \delta [p_T] \right )  &\equiv  \frac{1}{2}  \Delta \left( \left\langle  (\delta[p_T])^2 \right\rangle   \right), \\
\nonumber \Delta \left (  {\rm cov} (v_2^2, [p_T]) \right )  &\equiv  \frac{1}{3} \Delta \left( \left\langle  v_2^2 \delta[p_T] \right\rangle   \right), \\
\Delta \left ( v_2\{4\} \right )  &\equiv  \frac{1}{4} \Delta \left(  2 \left \langle v_2^2  \right\rangle^2 -  \left \langle v_2^4 \right \rangle    \right).
\end{align}
The proportionality factors in the relations $v_2 \propto \varepsilon_2$ and $[p_T] \propto E/S$ cancel out in the definition of $\Delta \mathcal{O}$. Therefore, the variations of the final-state observables in Eq.~(\ref{eq:obsVn}) can be estimated from the variations of the initial-state quantities in Eq.~(\ref{eq:obsEn}).

Our results are in Fig.\ref{fig:correlation_HIC_observables}(c)-(f) for ultra-central \ndnd{} collisions. We see that driven by the change in quadrupole deformation, both $v_2$ and the $[p_T]$ fluctuation [panel (c) and (d)] vary by several percent, in agreement with recent hydrodynamic studies \cite{Fortier:2023xxy,Fortier:2024yxs}. A similar variation is observed for the cumulant $v_2\{4\}$ [panel (f)], as nuclear deformation enhances the non-Gaussianity (kurtosis) of the distribution of the elliptic flow vector \cite{STAR:2015mki,Giacalone:2018apa,Mehrabpour:2023ign}. Finally, the covariance of $v_2^2$ and $[p_T]$ in Fig.~\ref{fig:correlation_HIC_observables}(e) presents the strongest variation, changing by about 10\%, with a positive correlation due to the fact that $v_2^2$ and $[p_T]$ are anti-correlated in the presence of a large $\beta_2$ \cite{Giacalone:2019pca,Giacalone:2020awm,Jia:2021wbq,Jia:2021qyu,Fortier:2023xxy,Fortier:2024yxs,STAR:2024wgy}.

Our main result is that all of these variations are strongly correlated with the value of the NME. The question, then, is whether the spread in NME values would be reduced if the nuclear model were constrained to ensure a precise description of the collective flow measurements following high-energy collision simulations. For this to be the case, the addition of high-energy data to the Bayesian analysis should effectively improve the extraction of the nuclear deformation, without introducing any substantial systematic uncertainties. We reiterate, then, that by combining flow observables among different collision systems of similar size (as done recently for \ruru{} vs. \zrzr{}, or \auau{} vs. \uuuu{}), it is possible to isolate signatures of the geometry of the colliding isotopes while suppressing the impact of the uncertainty on our knowledge of the QGP, such as the detailed values of its transport coefficient in hydrodynamics \cite{Xu:2021uar,Nijs:2021kvn,STAR:2024wgy,Mantysaari:2024uwn}. For example, taking the well-known near-spherical nucleus $^{208}$Pb as a baseline, the hydrodynamic studies suggest that a ratio of observables of the type: $
    \mathcal{O}\,[{\rm ^{150}Nd+^{150}Nd}]\,/\,\mathcal{O}\,[{\rm ^{208}Pb+^{208}Pb}]$, 
will depend almost exclusively on the differences in structure between $^{150}$Nd and $^{208}$Pb. Therefore,
we expect that a combined Bayesian analysis of low-energy nuclear structure data and high-energy heavy-ion data will lead to an improved determination of $\beta_2$. As a consequence, the uncertainty on the NME will decrease.
 
 \textit{\textbf{Summary and outlook.}} The results of a Bayesian analysis of low-energy nuclear structure data in the framework of CDFT are coupled to simulations of high-energy collisions, highlighting the existence of strong correlations between the NMEs of $0\nu\beta\beta$ decay and quantities accessible at colliders. Our conclusion is that it is possible to leverage measurements of the observables discussed in Fig.~\ref{fig:correlation_HIC_observables} to benchmark theoretical predictions of the NMEs, providing a complementary approach to low-energy nuclear physics methods.

This result has to be followed up by a combined Bayesian analysis of low- and high-energy data (likely by means of mock experimental data in the absence of collider results on the candidate isotopes), which would yield a better handle on the nuclear deformation and thus on the NME. In doing so, one should also include the octupole and the triaxial deformation of the candidate isotopes, which are expected to impact the values of the matrix elements, and which can be targeted by observables such as ${\rm cov}(v_2^2, [p_T])$, the triangular flow, $v_3$, or the skewness of $[p_t]$ fluctuations \cite{Bally:2021qys,Zhang:2021kxj,Jia:2021qyu,Nielsen:2023znu}. Moreover, it should be investigated whether combining data sets from \ndnd{} and \smsm{} collisions (or other pairs of candidates) gives access to \textit{relative observables} that may present an even tighter correlation with the NME.

We note that, from a high-energy experiment point of view, it is rather straightforward to collect a statistics of ultra-central events sufficient to measure observables including $v_2\{4\}$ with a relative uncertainty of 1\% or less, which seems required from the plots shown in Fig.~\ref{fig:correlation_HIC_observables}. At the Large Hadron Collider (LHC), and thanks to upcoming detector upgrades such as ALICE 3 \cite{ALICE:2022wwr}, such results could be obtained via short special collision runs of only a few hours. The potential upgrade of the ion injector complex with the addition of a second ion source \cite{AlemanyFernandez:2025ixd} will enable us, furthermore, to collide ions in the aforementioned isobar operation mode, which will reduce experimental and theoretical systematic uncertainties on the flow measurements, isolating clean signals of the nuclear geometries. Therefore, a potential spin-off of the LHC ion program connecting with the next generation of $0\nu\beta\beta$ decay searches appears technically within reach and worth pursuing.

It is worth noting that the present study is based on MR-CDFT. Extending this analysis to cutting-edge nuclear \textit{ab initio} methods, such as the in-medium generator coordinate method~\cite{Yao:2020PRL}, would be highly interesting and will be discussed in a future publication. Moreover, as previous studies~\cite{Rodriguez:2004,Simkovic:2004NPA,Yousef:2009,Pacearescu:2004} have demonstrated, the NMEs of single-beta decay and two-neutrino double-beta decay are also strongly affected by nuclear deformations, such that it would be interesting to extend the present study to these processes too.

\begin{acknowledgments}
\textit{\textbf{Acknowledgments.}}
We thank Lotta Jokiniemi, Huichao Song, as well as the participants of the CERN workshop \textit{``Nuclear Shape and BSM Searches at Colliders''} for useful discussions. This work was supported in part by the National Natural Science Foundation of China (Grant Nos. 12375119 and 12141501), and the Guangdong Basic and Applied Basic Research Foundation (2023A1515010936). 
 
\end{acknowledgments}


%

\end{document}